\documentclass{aa}

\usepackage{natbib}
\bibpunct{(}{)}{;}{a}{}{,}
\usepackage{graphicx}
\usepackage{txfonts}
\usepackage{lipsum}
\usepackage{xcolor}
\usepackage{amsmath}
\usepackage{caption}
\usepackage{subcaption}
\usepackage{lscape}
\usepackage{placeins}

\usepackage[]{hyperref}

\begin{document}

   \title{Rotational equilibrium of $\mathrm{C}_2$ in diffuse interstellar clouds}

   \subtitle{II - Formation-excitation at work? }

   \author{J. Le Bourlot\inst{1}\fnmsep\inst{2}\fnmsep\thanks{38 years have passed since the first paper of this "series" \citep{1987A&A...188..137L}. Never give up hope...}
          \and
          E. Roueff\inst{1}
          \and
          S. R. Federman\inst{3}
          \and
          A.M. Ritchey\inst{3}
          \and
          D. L. Lambert\inst{4}
          }

   \institute{LUX, Observatoire de Paris, Université PSL, Sorbonne Université, CNRS, 92190 Meudon, France\\
              \email{Jacques.Lebourlot@obspm.fr}
         \and
         Université Paris-Cité
         \and
             University of Toledo, Department of Physics and Astronomy, Toledo, OH 43606, USA\\
             \email{steven.federman@utoledo.edu}
        \and
             University of Texas at Austin, Department of Astronomy, Austin, TX 78712, USA\\
             \thanks{Based on observations obtained with the Hobby-Eberly Telescope (HET), which is a joint project of the University of Texas at Austin, the Pennsylvania State University, Ludwig-Maximillians-Universitaet Muenchen, and Georg-August Universitaet Goettingen. The HET is named in honor of its principal benefactors, William P. Hobby and Robert E. Eberly.}
           }

   \date{Received; accepted }

  \abstract
  {Recent spectroscopic measurements have revealed absorption from higher rotational levels in $\mathrm{C}_2$ than previous observations. These improvements are accompanied by the availability of updated radiative and collisional data.}
   {We revisit the density and radiation field intensity diagnostics provided by the observations of many rotational levels of interstellar $\mathrm{C}_2$ and extensive molecular information.}
   {We built an excitation model of $\mathrm{C}_2$ without spatial structure, including levels up to $J$ = 34 where updated radiative and collisional excitation data are introduced as well as excitation by chemical formation.}
 {We confirm the importance of the recent collisional excitation rate coefficients of $\mathrm{C}_2$ by molecular $\mathrm{H}_2$. We show that the new higher level observations cannot be explained by the standard balance between collisional excitation and radiative transitions. We propose that chemical excitation at formation provides a plausible mechanism to explain the observed high excitation of $\mathrm{C}_2$. In addition, it allows us to lift the degeneracy of the density over radiation field strength parameter in the excitation model.}
   {A 0D model remains limited and it is highly desirable to use a full Photon Dominated Region (PDR) model, which includes all excitation processes introduced here and full chemical and thermal balance.}


   \maketitle

\section{Introduction}

The first detection of diatomic carbon by \citet{1977ApJ...216L..49S} towards Cygnus OB2-12 was obtained in the near infrared (1-0) Phillips band. That study was followed by several other detections in the same electronic system.
\citet{1995ApJ...438..740L} made definitive interstellar detections of the $D-X$ and $F-X$ electronic transitions from ultraviolet observations with the Hubble Space Telescope. \citet{2007ApJS..168...58S} and \citet{2012ApJ...761...38H} expanded the sample of sight lines with UV measurements.
 These observations allow us to infer the column density of the different rotational levels up to relatively high values of $J$ and are thus promising diagnostic tools of the physical conditions of the associated environments if the different excitation mechanisms are well accounted for. Since then, higher resolution observations have been obtained in the visible and near infrared with the Akari telescope \citep{2019ApJ...881..143H}, the very large telescope ultraviolet and visual echelle spectrograph (VLT UVES) \citep{2024A&A...681A...6F}, and the new habitable zone planet finder (HPF) (present work).
 The Hobby-Eberly telescope (HET) HPF spectra significantly extend observations of the $v=0$ $X$ state rotational ladder beyond previous upper limits, which opens the possibility of novel tests.

\citet{1982ApJ...258..533VC} presented the first detailed analysis of $\mathrm{C}_2$ excitation by computing the quadrupole radiative transition probabilities within the ground state up to $J=16$, and the intercombination transitions between levels of the lowest triplet and singlet states, as well as estimating the rotational collisional de-excitation rates due to collisions with $\mathrm{H}$ and $\mathrm{H}_2$. \citet{1986JMoSp.120..157L} improved the treatment of the intercombination transitions and introduced them into a model applied to the $\zeta$ Oph line of sight.

More recently, \citet{2012ApJ...749...48C} included new spectroscopic results on the $\mathrm{C}_2$ Phillips band and collisional excitation rates due to $\mathrm{He}$, and proposed
 a two-phase model that enabled them to explain the observations in a few diffuse and translucent lines of sight. Finally, \citet{2024ApJ...973..143N} introduce the new collisional rate coefficients for the excitation of $\mathrm{C}_2$ due to para- and ortho-molecular hydrogen \citep{C9RA10319H} by using the spectroscopic information reported in \citet{2012ApJ...749...48C}. Their study leads to a significant reduction in the density estimates as a consequence of the larger collisional rate coefficients computed by \citet{C9RA10319H}.

In this paper, we revisit the excitation mechanisms pertaining to interstellar
 \footnote{A simplified term diagram of the $\mathrm{C}_2$ molecule is shown on Fig.\,\ref{fig:Electr}} $\mathrm{C}_2$ and present a 0D excitation model that is further compared to new observations.
 In Sect.\,\ref{Excitation} we describe the detailed balance equations and the different processes that contribute to the population of $\mathrm{C}_2$ rotational levels, including collisional, radiative, and chemical formation contributions.
Sect.\,\ref{Synthesis} details the importance of the various contributions for a
given set of temperatures and densities. In Sect.\,\ref{Observations} we report the observations of $\mathrm{C}_2$ in Cyg OB2-12, HD29647 performed with the HPF and in HD24534 \citep{2007ApJS..168...58S} and in Sect.\,\ref{Comparison} we present our derivation of physical conditions through an optimization schema of our excitation model. Our conclusions are summarized in Sect.\,\ref{Discussion}.

\section{Excitation mechanisms of $\mathrm{C}_2$}\label{Excitation}
\subsection{Detailed balance equations}
It is customary to study the detailed balance of a species by only considering collisions and radiative transitions within the set of levels considered. However, this leads to serious inaccuracies in the resulting populations if other physical processes occur with characteristic times comparable to collision times or radiative lifetimes. In the case of $\mathrm{H}_3^+$, \citet{2024MolPh.12282612L} have shown that chemical destruction by electrons may be fast enough to prevent thermalization of the lowest ortho level of the molecule, leading to an excitation temperature between the two lowest levels that is typically half that of the gas kinetic temperature.

These state-specific chemical rates are easily included in the balance equation and have the advantage, for the theoretician, of providing a solution without using mass conservation. Hence we have the further benefit of a direct check of the quality of the solution.
The general form of the balance equation for a level $i$ of any species is thus
\begin{equation}
    x_{i}\,\left(\sum_{j}R_{ij}+D_{i}\right)=\sum_{j}x_{j}\,R_{ji}+F_{i}
    \label{Eq:Steady_state}
.\end{equation}
Here, $x_i$ is the relative abundance of level $i$, $R_{ij}$ is a rate of transition (including inelastic collisions and radiative decay or pumping) from level $i$ to level $j$ (respectively from $j$ to $i$) in $\mathrm{s}^{-1}$, $D_i$ is the destruction rate of $\mathrm{C}_2$ on level $i$ by "chemical" processes in $\mathrm{s}^{-1}$, and $F_i$ is the formation rate of level $i$ in $\mathrm{s}^{-1}$.
It is important to note that excitation can never be decoupled from other physical processes. State-specific formation and destruction must always be accounted for. These processes are included here for the first time.

The following subsections detail the different contributions to $R_{ij}$. The electronic $X\,^1\Sigma_g^+$ ground state of $\mathrm{C}_2$ is very close to the $a\,^3\Pi_u$ excited state so that to compute accurate ground-state populations of rotational levels up to $J = 32$, as found towards some lines of sight, one must also include rotational levels belonging to $a\,^3\Pi_u$ that fall between rotational levels of $X$. In addition, vibrationally excited levels with a low rotational quantum number belonging to $X$ may decay towards $a$\footnote{The $X-a$ radiative transition probabilities are much larger than those involving vibrational electric quadrupole decay, despite the spin flip.}. Finally, we computed in a consistent way the detailed balance of the first $120$ levels that encompass the $v=0$, $J=34$ level of $X$, in order to avoid border effects.

\begin{figure}
\centering\includegraphics[width=1.0\columnwidth]{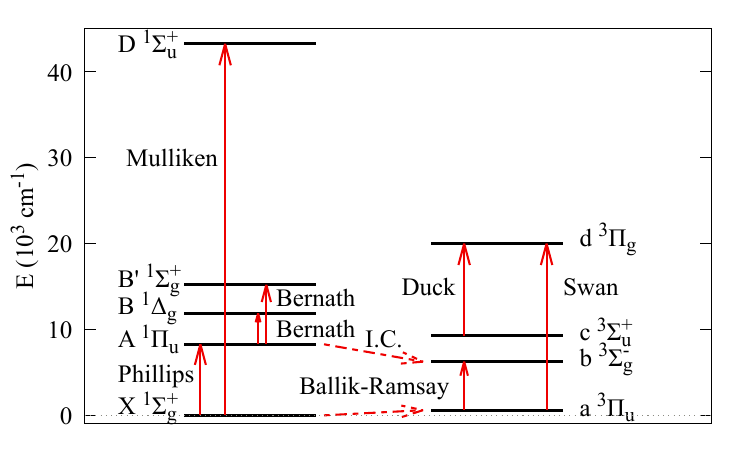}

\includegraphics[width=1.0\columnwidth]{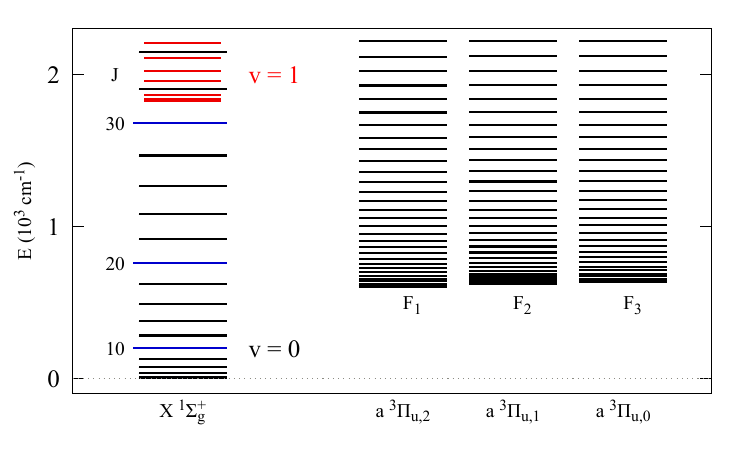}
\caption{Top: $\mathrm{C}_2$ electronic systems included in this computation. "I.C." stands for "Intercombination" (inspired by \citet{2018MNRAS.480.3397Y}). Bottom: Ro-vibrational levels included in the full detailed balance computation. Numerical values are given in Table\,\ref{tab:Energy_Levels}}\protect\label{fig:Electr}
\end{figure}

The top panel of Fig.\,\ref{fig:Electr} shows the electronic states included in this computation through radiative pumping followed by cascades (see Sect.\,\ref{Sec:Cascades}); the bottom panels shows the position of all $120$ levels included in the full computation. This includes $18$ rotational levels within $X$, $v=0$, $8$ rotational levels within $X$, $v=1$, and $94$ levels within $a$. One can see that the density of levels is much higher in the $a$ state, which leads to a dilution of abundances.

\subsection{Collisions}

\begin{figure}
\centering\includegraphics[width=1.0\columnwidth]{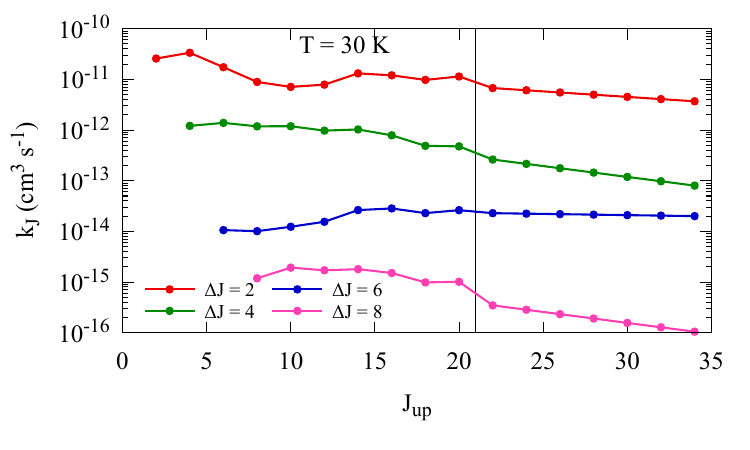}

\includegraphics[width=1.0\columnwidth]{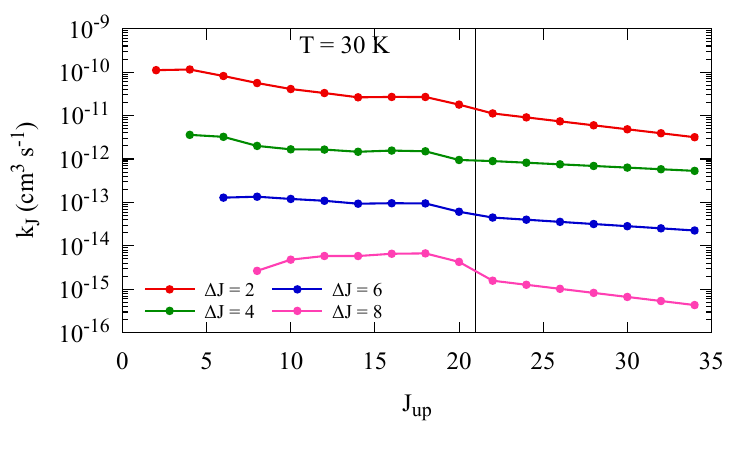}
\caption{Extrapolation of collisional rate coefficients from \citet{2008CPL...460...31N} at $30\,\mathrm{K}$.
Extrapolated rate coefficients are on the right side of the black
vertical line. Top: Para-$\mathrm{H}_{2}$, Bottom: Ortho-$\mathrm{H}_{2}$.\protect\label{fig:Extrapolation-of-collision}}
\end{figure}

The $\mathrm{C}_2$ molecule forms mainly in the molecular part of clouds. Hence, the main collision partner by far is $\mathrm{H}_2$.
\citet{C9RA10319H} provide separate rate coefficients for para- and ortho-$\mathrm{H}_2$ collisions within $X$ for $J \leqslant 20$. This is not enough to compute accurate populations up to $J = 32$. Thus, we have extrapolated these rates for $\Delta J = 2$, $4$, $6$, and $8$ with a simple exponential fit. Although rather crude, this approximation is better than setting the rates to $0$. It is shown on Fig.\,\ref{fig:Extrapolation-of-collision} and numerical values are provided in Tables\,\ref{tab:Ortho-H2-Col} and \ref{tab:Para-H2-Col}.

Collisional excitation rate coefficients due to $\mathrm{He}$ are found in \citet{2008CPL...460...31N} (Table\,1) for $T = 5$ to $100\, \mathrm{K}$ up to $J = 20$ and in \citet{RITIKA2022139623} for very low temperatures and a few $J$. They agree roughly at $5\,\mathrm{K}$. The same procedure as for collisional rates with $\mathrm{H}_2$ was used to extend the available rates to higher $J$ for $\Delta J = 2$ and $4$, as found in Table\,\ref{tab:He-Col}.

To the best of our knowledge, no collision rates exist for collisions with $\mathrm{H}$, so all models below are computed for a fully molecular gas. If collision rates with $\mathrm{H}$ are similar to those with para-$\mathrm{H}_2$ (as is often the case), then a higher density $n_\mathrm{H}$ would be required to induce the same impact of collisions on excitation.
For collision-induced transitions, $R_{ij} = k^{\mathrm{X}}_{ij} \, n_{\mathrm{X}}$, where $k^{\mathrm{X}}_{ij}$ is the rate coefficient for collisions with species $\mathrm{X}$ of abundance $n_{\mathrm{X}}$.

\subsection{Electric quadrupolar radiative transitions within $X$ ground electronic state}

\begin{figure}
\centering\includegraphics[width=1\columnwidth]{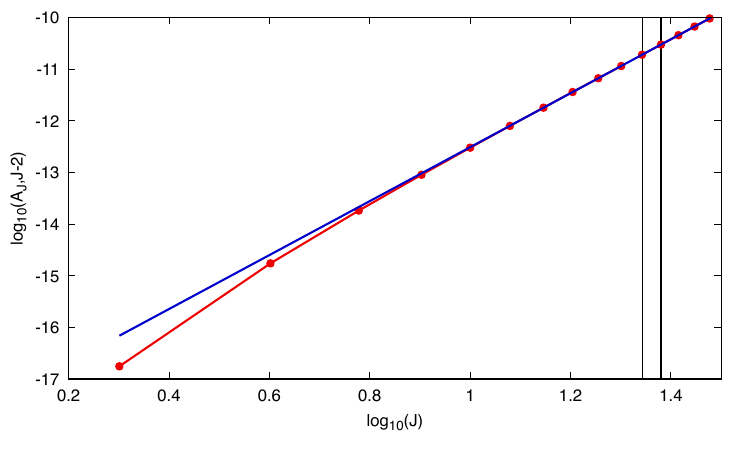}
\caption{Electric quadrupole radiative rotational transition probabilities from \citet{1982ApJ...258..533V}.
The positions of $J=22$ and $24$ are shown by straight lines. The points
above are extrapolations.\protect\label{fig:_C2_Quad}}
\end{figure}

As $\mathrm{C}_2$ is a homonuclear molecule, there are no electric dipolar transitions within the $X$ electronic state. Quadrupolar transition probabilities have been computed by \citet{1982ApJ...258..533V} but only up to $J = 20$ (their Table\,2). Here, we propose to use the energy dependence of these expressions to justify an extrapolation scheme. Electric quadrupole emission coefficients for $J'\rightarrow J"$ are given by
\begin{equation}
    A_{J',J"}=\frac{32\,\pi^{6}}{5\,h}\, \bar{\nu}^{5}\,\frac{S_{Q}\left(J',J"\right)}{2J'+1}
,\end{equation}
where $\bar{\nu}$ is the transition wavenumber in $\mathrm{cm}^{-1}$ and the strength $S_{Q}\left(J',J"\right)$ is in atomic units:
\begin{equation}
S_{Q}\left(J',J"\right)=\left|\int\psi_{J'}^{*}\,Q\,\psi_{J"}\,dR\right|^{2}\,\frac{J'\,\left(J'-1\right)}{2J'-1} \label{Eq:S_Q}
\end{equation}
$\bar{\nu}\left( J' \rightarrow J'-2\right) = 4\,B\,J'- 2\,B$, with $B$ being the rotational constant of the molecule, so that $\bar{\nu}^{5} \propto J^5$. From Eq.\ref{Eq:S_Q} we see that, for high rotational $J$ values, $\frac{S_{Q}\left(J',J"\right)}{2J'+1}$ does not depend on $J$. So $A_{J',J"}$ varies as $J^5$ (neglecting the
possible variation of the electric quadrupolar transition matrix
element). Plotting $\log_{10}\left(A_{J',J"}\right)$ as a function of $\log_{10}(J)$, we see that the points corresponding to the highest $J$ transitions are aligned on a straight line (Fig.\,\ref{fig:_C2_Quad}). The transition probability values extrapolated from a linear fit and the transition wavenumbers obtained from the ExoMol database\footnote{\href{https://www.ExoMol.com/data/molecules/C2/12C2/8states/}{https://www.ExoMol.com/}.} are reported
in Table\,\ref{tab:Quadrupolar-transitions-proba}.

For $v>0$ transitions, \citet{1982ApJ...258..533V} only give the sum $A\left(v',J'\right)$ over all lower levels, and state that these $A\left(v,J\right)$ are "virtually independent of $J$". Using the expressions of $S\left(J',J"\right)$ for the $S$, $Q$, and $O$ branches, we can distribute these sums to recover individual $A_{v'=1J',v"=0J"}$, as displayed in Table\,\ref{tab:v_1-0_Aul}.
For electric quadrupolar transitions, $R_{ij} = A_{ij}$, the Einstein emission coefficient.

\subsection{Electronic radiative processes}\label{Sec:Cascades}

In their Fig.\,2, \cite{1982ApJ...258..533V} describe the different radiative electronic transitions that link the rotational levels of the $X\,^1\Sigma^+_g$ ground electronic state. However, in the absence of complete radiative information, these authors introduced some approximations to account for these mechanisms.
Radiative electronic transitions data within the eight lowest electronic states of $\mathrm{C}_2$ are now accessible and reported in the ExoMol database, extracted from the recent study of \citet{2020MNRAS.497.1081M}. The database includes transitions within singlet states $X\,^1\Sigma^+_g$, $A\,^1\Pi_u$, $B\,^1\Delta_g$, and $B'\,^1\Sigma^+_g$ and within triplet states $a\,^3\Pi_u$, $b\,^3\Sigma^+_g$, $c\,^3\Sigma^-_u$, and $d\,^3\Pi_g$. The ExoMol database also incorporates intercombination transitions between singlet and triplet states. We have supplemented it with transitions to $D\,^1\Sigma^+_u$ (Mulliken system) with data from \cite{2007JChPh.127w4310S} (oscillator strengths) and \citet{2018JMoSp.344....1K} (energy level positions and derived transition wavenumbers). This leads to a set of over $40000$ levels coupled by more than $600000$ lines. Within this total of $N=44189$ levels, we only computed a detailed balance for the lowest $M$ ones ($M = 120$ here to include all levels up to $X$, $v=0$, $J=34$, just above the highest level observed; see Sect.\ref{Sect:Cyg-OB2}).

These data provide the spontaneous radiative emission coefficients $A_{ij}$ and corresponding transition wavenumbers $\bar{\nu}$, from which absorption $B_{ji}$ and induced emission $B_{ij}$ coefficients can be derived.
In the presence of an external radiation field, such as the ambient interstellar radiation field (ISRF), absorption toward electronically excited states is possible, followed by radiative cascades to a low lying level.
This mechanism results in the possible coupling of two low lying levels even if no direct transition is possible between them. For two levels $i$ and $j$ below $M$, the coupling coefficient by this indirect cascade, $R_{ij}$, can be written in the optically thin approximation as
\begin{equation}
    R_{ij} = \sum_{n>M} B_{in} \, \bar{J}_{in} \, C_{nj}
,\end{equation}
where $B_{in}$ is the Einstein absorption coefficient to a level $n$ above $M$, $\bar{J}_{in}$ is the mean radiation field intensity, and $C_{nj}$ is the cascade coefficient from $n$ to $j$. For a fixed value of $M$, the coefficients $C_{nj}$ may be computed in advance and stored. With $M=120$, we end up with $N=12878$ levels, which are accessible through $71708$ radiative transitions and lead to $1463887$ cascade coefficients. The algorithm to compute the cascade coefficients is given in Appendix\,\ref{Cascade_Coef}.

\subsection{Adopted ISRF}

\begin{figure}
\centering\includegraphics[width=1\columnwidth]{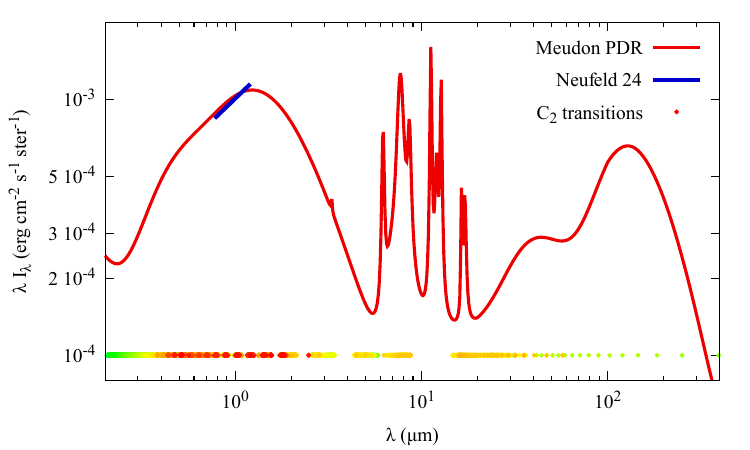}

\caption{Adopted ISRF for $G=1$ in the range of possible transitions of $\mathrm{C}_2$. Colored points displayed on the horizontal line at the bottom of the figure indicate the positions of included transitions coded with the value of $B_{lu}\,J_{ul}$ from red (strongest) to light green (weakest).}
\label{Fig:ISRF}
\end{figure}

The adopted ISRF is the one used in the Meudon PDR code\footnote{Available at \href{https://pdr.obspm.fr}{https://pdr.obspm.fr}.}, based on \citet{1983A&A...128..212M}. It is defined as a reference and corresponds to G=1.
That dimensionless factor will be used further for possible future scalings of the ISRF. However, the scaling does not apply in the millimeter domain where the Cosmic Microwave Background (CMB) contribution applies.
It is displayed in Fig.\,\ref{Fig:ISRF}. It extends from the ultraviolet (where transitions from the Mulliken system take place) to the submillimeter range for the lowest rotational transitions.
\citet{2024ApJ...973..143N} only consider the $0.77 - 1.21\, \mu\mathrm{m}$ restricted spectral window, covering the Phillips $A\,^1\Pi_u$ - $X\,^1\Sigma_g$ band system of $\mathrm{C}_2$,
as shown on top of the present adopted curve in Fig.\,\ref{Fig:ISRF},
 where an almost perfect agreement is found with our adopted ISRF in
 the considered spectral range.

However, additional noticeable absorptions take place outside that window. Horizontal points on Fig.\,\ref{Fig:ISRF} show the positions of these absorptions and are color-coded proportionally to $B_{lu}\,J_{ul}$. In particular, weak but non-negligible absorptions also occurs in the mid-infrared region at the positions of some polycyclic aromatic hydrocarbons (PAH) features.

We computed the resulting absorption probability of the $X$ and $a$ levels included in our excitation model for $G = 1$ and found an almost $J$ independent value of $4.3\,10^{-9}\, \mathrm{s}^{-1}$ for the rotational $X$ levels and obtained a value of $\sim 6.7\, 10^{-9} \, \mathrm{s}^{-1}$ for the $a$ levels.
We note that \cite{2024ApJ...973..143N} mention this same rotational independent value for their estimated pumping rate of the $X$ rotational levels that results from their update of the radiation field used previously by \cite{1982ApJ...258..533V}.

\subsection{Chemical excitation}\label{Sec:Chem_Exc}

In the cold ISM, $\mathrm{C}_2$ is mostly formed by exothermic reactions. Detailed examination of the various formation paths shows that no single reaction dominates, but that a few contribute at the $10$ to $20\,\%$ level. Among these, some reactions have a large exothermicity. The main ones are:
\begin{eqnarray*}
   \mathrm{C_2H}^+ + e^- & \rightarrow & \mathrm{C}_2 + \mathrm{H} \quad (\Delta H = 6.671 \, \mathrm{eV}),\\
   \mathrm{C_3}^+ + e^- & \rightarrow & \mathrm{C}_2 + \mathrm{C} \quad (\Delta H = 4.804 \, \mathrm{eV}),\\
   \mathrm{C}^+ + \mathrm{CH} & \rightarrow & \mathrm{C}_2 + \mathrm{H}^+ \quad (\Delta H = 0.321 \, \mathrm{eV}).
\end{eqnarray*}
Electronic recombination rates are found in \citet{2006PhR...430..277F}. Branching ratios for light carbonaceous molecules are found in \citet{2013ApJ...771...90C}.
Level number $120$ of $\mathrm{C}_2$, corresponding to $F1$, $a\,^3\Pi_u,\, v=1,\, J=2$, is about $3000\,\mathrm{K}$ above the ground level. So, if a significant fraction of the exothermicity remains in the molecule, it is enough to populate levels proportionally to their statistical weight. This is a zeroth order approximation, which allows for easy testing of the importance of chemical formation excitation, compatible with the current state of knowledge.\footnote{The authors thank Octavio Roncero, Steve Kable, and Chris Hansen for insightful discussions on $\mathrm{C}_2$ internal excitation at formation.}

In the following, we take
\begin{equation}
    F_i = k_f \, \frac{g_i}{\sum_M g_j}
,\end{equation}
where $k_f$ (in $\mathrm{s}^{-1}$) is a single effective parameter that accounts for all chemical reactions leading to internal excitation and $g_i$ is the statistical weight of level $i$ ($f$ is for "formation"). All $M$ levels are included, both from $X$ and $a$. We define $k_f$ by
\begin{equation}
    k_f = \frac{1}{[\mathrm{C}_2]}\, \left. \frac{d[\mathrm{C}_2]}{dt} \right|_f
    \label{Eq:k_f}
.\end{equation}
Optimal values for $k_f$ derived below have been checked for consistency with the results from the Meudon PDR code, where the full chemistry is solved, including over $8000$ reactions between over $230$ species. Typical values go from a few $10^{-13}$ to $10^{-11}\,\mathrm{s}^{-1}$, depending on cloud structures and conditions.

No simple reasoning leads to a particular choice of state-specific destruction rates. So we take simply $D_i = k_f$ (in $\mathrm{s}^{-1}$), which ensures that
\begin{equation}
    \sum_i x_i \, D_i = \sum_i F_i.
\end{equation}

\section{Results from 0D model}
We now discuss the numerical results{\footnote{The code \texttt{ExcitC2} developed for this work is available upon request, and it should be available soon on the \href{https://ism.obspm.fr}{https://ism.obspm.fr} web site.} obtained after including the various excitation
processes in the balance equations (Eq.\ref{Eq:Steady_state}). The input parameters are the density, $n_\mathrm{H}$, the temperature, $T$, the ambient radiation field scaling, $G$, and the chemical formation rate, $k_f$. The output values are the relative populations of $\mathrm{C}_2$.

The system of $M$ equations is linear and well behaved, and no conservation equation is needed if $k_f \neq 0$. It is easily solved using, e.g., the \texttt{LAPACK} library. However, using $120$ levels leads to a system with a rather high condition number and requires the use of the \texttt{DGESVX} extended routine to allow for matrix equilibration.
We suppose that hydrogen is fully molecular, so that $n\left( \mathrm{H}_2 \right) = n_\mathrm{H} / 2$, and that the ortho to para ratio of $\mathrm{H}_2$ is at thermal equilibrium. This assumption is fine for diffuse cloud conditions where the gas temperature is derived from the $J$=1 to $J$=0
ratio of $\mathrm{H}_2$ \citep{2021ApJ...911...55S,2024MolPh.12282612L}. We also include helium, which represents 20\% of the molecular $\mathrm{H}_2$ density. Here we emphasize the very large dependency of the collisional excitation rate coefficients of $\mathrm{C}_2$ on the ortho to para ratio of $\mathrm{H}_2$, as already pointed out by \cite{2024ApJ...973..143N}.
Such a 0D model is restricted to study the relative populations of a defined number of $\mathrm{C}_2$ levels without considering the detailed chemical or thermal balance and independent of geometry.
It may be considered as an extension and update to the Ben McCall $\mathrm{C}_2$ Calculator that includes only seven levels of $\mathrm{C}_2$ \footnote{available at https://dib.uchicago.edu/c2/.}.

\subsection{A fiducial example}\label{Synthesis}

\begin{figure}
\centering\includegraphics[width=1\columnwidth]{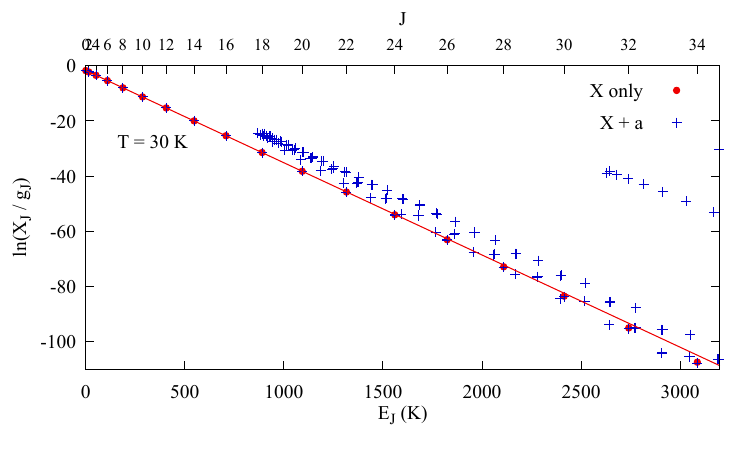}
\caption{$\mathrm{C}_2$ excitation diagram for $T = 30\,\mathrm{K}$ and $n_\mathrm{H} = 50\,\mathrm{cm}^{-3}$ without radiative cascades. Red: Only levels from $X$ included (labeled "$X$ only"); Blue: $X$ and $a$ electronic states, including intercombination transitions (labeled "$X + a$").\protect\label{fig:Base_PureX}}
\end{figure}

\begin{figure}
\centering\includegraphics[width=1\columnwidth]{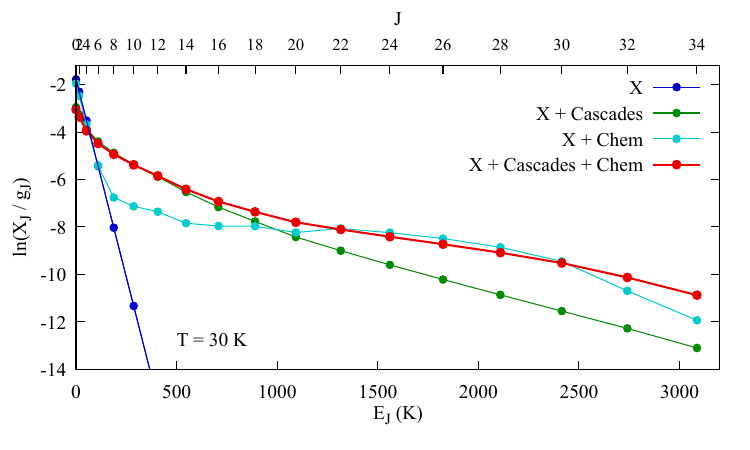}
\caption{Effect of nonthermal excitation mechanisms on the weighted fractional populations (see text for the explanation of the labels).\protect\label{fig:Base}}
\end{figure}

The impact of these various excitation mechanisms is best seen on a representative example. We consider gas at a temperature of $T = 30\,\mathrm{K}$ and $n_\mathrm{H} = 50\,\mathrm{cm}^{-3}$ and introduce the different excitation mechanisms step by step. In the following, we assume a radiation field scaled by the value $G=1$. We first examine the dependence of the weighted fractional abundances of $\mathrm{C}_2$ on the inclusion of state $a$. Only collisions and direct radiative absorption or emission are accounted for (Fig.\,\ref{fig:Base_PureX}).
Using only the rotational levels from $X$ (red points), we recover a pure Boltzmann distribution at the gas temperature. This is because quadrupolar transitions are weak and even a density of $100\,\mathrm{cm}^{-3}$ is above the critical density.
Adding the $a$ levels, which are in the same energy range as the rotational levels belonging to $X$ (the previously mentioned 120 levels), leads to the weighted populations in blue. Radiative pumping from low-lying rotational levels of $X$ gives a perceptible population of state $a$ showing up at an energy of $\sim 800\,\mathrm{K}$. The fractional populations within $X$ remain the same.

Fig.\,\ref{fig:Base} is restricted to the X rotational levels, although levels from $a$ are included in the computation. This shows the effect of adding further excitation mechanisms. Blue points (label $X$) are the same as in Fig.\,\ref{fig:Base_PureX}. Green points show the effect of radiative pumping from Sect.\,\ref{Sec:Cascades} (label "$X$ + Cascade"). This approximation represents a
significant update from the previous treatments \citep{1982ApJ...258..533V,2012ApJ...749...48C,2024ApJ...973..143N}, where intercombination transitions are included approximately over a restricted spectral range. The corresponding weighted fractional abundances of $\mathrm{C}_2$ decrease smoothly as a function of the rotational energy.
We have also tested this case with and without the inclusion of the Mulliken electronic band system in the cascade coefficients. The differences are insignificant up to $J = 30$ and increase to only $10\,\%$ for $J = 34$.

Light blue points refer to a computation without considering radiative cascades
but with chemical pumping (Sect.\,\ref{Sec:Chem_Exc}) with $k_f = 8\,10^{-12}\,\mathrm{s}^{-1}$ (label "$X$ + Chem"). Chemical pumping maintains an almost constant abundance for $J > 6$. Red points include all contributions
together (label "$X$ + Cascade + Chem"). Radiative cascades dominate for intermediate $J$ (from $J = 6$ to $12$); then chemical excitation maintains
a slowly decreasing population.

\subsection{Parameter dependencies}\label{Sec:param_dep}
Such a model is hopefully useful to extract the relevant physical conditions from
the excitation properties. In the following, we discuss the dependence of these on density, temperature, and radiation field.
We consider chemical pumping effects that impact only the most rotationally excited levels separately.

\begin{figure}
\centering\includegraphics[width=1\columnwidth]{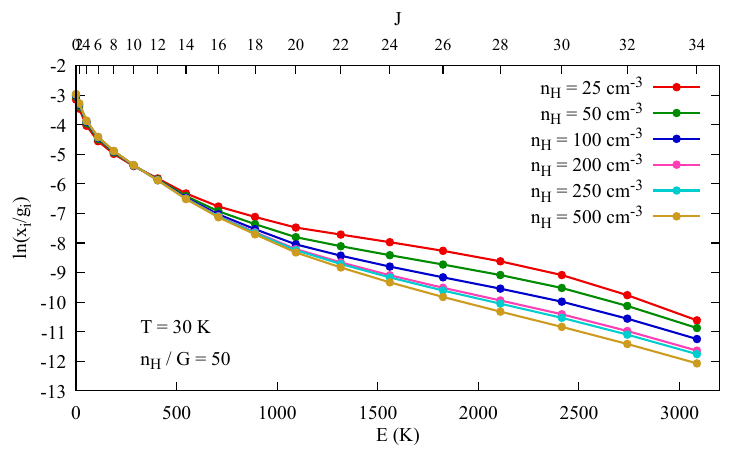}
\caption{Excitation diagram for a ratio $n_\mathrm{H}/G = 50$. All results collapse on a single curve up to $J = 16$. Note that the density range is much larger than in the cases of Appendix \ref{Variations}}
\protect\label{fig:Pum=1}
\end{figure}

Fig.\,\ref{fig:Pum=1} shows an excitation diagram computed for $T = 30\,\mathrm{K}$, $k_f = 0$ and a range of densities $n_\mathrm{H}$. However, we have set a fixed value to the ratio
$\alpha = n_\mathrm{H}/G$. We see that all curves collapse on a single excitation diagram up to level $J = 16$. This may be understood by considering the various terms of Eq.\ref{Eq:Steady_state}. We observe that quadrupolar spontaneous transitions $A_{ij}$ are much smaller than other terms for $J < 16$, and can be neglected. If we
additionally neglect chemical terms $D_i$ and $F_i$, then Eq.\ref{Eq:Steady_state} reduces to
\begin{equation}
n_{i}\,\sum_{j}\left(k_{ij}^{X}\,n_{X}+\sum_{n>M}B_{in}\,\bar{J}_{in}\,C_{nj}\right) = \sum_{j}n_{j}\,\left(k_{ji}^{X}\,n_{X}+\sum_{n>M}B_{jn}\,\bar{J}_{jn}\,C_{ni}\right)
.\end{equation}

The density of collision partners $n_X$ is proportional to $n_\mathrm{H}$, while the mean radiation field $\bar{J}$ is proportional to $G$. Thus, substitution of $n_\mathrm{H}$ by $\alpha\,G$ allows for a cancellation by $G$.
The resulting populations only depend on $\alpha$.
Overcoming this degeneracy requires additional information. The presence of nearby stars, in particular, may impact the strength and the shape of the radiation field.
The resulting population departs from this scaling if one of the neglected terms starts to play a role. This is the case for high $J$ levels, for which the emission coefficients $A_{ij}$ are large enough to contribute, as seen in Fig.\,\ref{fig:Pum=1} (see Table\,\ref{tab:Quadrupolar-transitions-proba}). Including chemical pumping ($k_f \neq 0$) also lifts the degeneracy, since $k_f$ does not scale with $n_\mathrm{H}$ (See Eq.\,\ref{Eq:k_f}).
As seen below, its effect on high $J$ levels is much more efficient than radiative pumping and cascades alone.
We have checked that this result remains true for all relevant temperatures (between $10$ and $100\,\mathrm{K}$).

In Appendix\,\ref{Variations}, we study the sensitivity of the excitation diagram to systematic individual variations of the physical parameters, temperature, density, and strength of the radiation field. It is clearly shown that none of those enables us to
obtain enhanced populations for $J$ values larger than $\sim 18$, as found in the observations. Then, chemical excitation appears as the only mechanism that allows such a nearly constant ratio $N_J/g_J$, as observed for high $J$s.

\section{Observations}\label{Observations}

We consider three different lines of sight in order to appraise our excitation model against observations.
Observations toward HD~29647 and Cyg~OB2~12 were acquired with the HPF on the 10 m HET at the McDonald Observatory \citep{2014SPIE.9147E..1GM,10.1117/12.2313835} as part of a larger sample to examine $\mathrm{C}_2$ and $\mathrm{CN}$ excitation and isotopologue ratios at near-infrared wavelengths. Here we provide a brief summary of the data acquisition and processing, leaving the detailed description to a future paper (Federman et al., in prep.). Besides these targets, spectra of telluric standards were obtained to remove lines arising from Earth's atmosphere. The $\mathrm{C}_2$ lines from the (2-0) and (1-0) bands of the Phillips system were used in the present analyses. The detector was a 2048x2048 pixel Hawaii-2 HgCdTe device with a resolving power of about $53,000$. The software package \emph{Goldilocks
\footnote{\href{https://github.com/grzeimann/Goldilocks\_Documentation}{https://
github.com/grzeimann/Goldilocks\_Documentation}}} was used for data reduction. The extracted one-dimensional spectra were converted to air wavelengths through the use of Eq. 16 in \citet{2003ApJS..149..205M}. After dividing the target spectrum by that of the telluric standard, the target spectrum was normalized and placed on the local standard of rest. The final spectra had signal-to-noise ratios per resolution element of about 2500.

The $\mathrm{C}_2$ spectra from HPF were fit using the profile synthesis code $ismod$ \citep{2008ApJ...687.1075S}, from which the velocity ($v_{\mathrm{LSR}}$), column density for each rotational level ($N(J)$), and Doppler parameter ($b$-value) were obtained.
We used the experimental measurements on $\mathrm{C}_2$ wavelengths from \citet{1977JMoSp..68..399C} and the results of theoretical results for oscillator strength by \citet{2007JChPh.127w4310S}.
For the gas toward HD~29647 and Cyg~OB2~12, as starting points we adopted the previously determined component structures by \citet{2021ApJ...914...59F} and \citet{2019ApJ...881..143H}, respectively.
Our results for HD 29647 are generally consistent with earlier measurements \citep{2021ApJ...914...59F}, and the agreement between our results for Cyg\_OB~12 and those of \citet{2019ApJ...881..143H} is very good (Federman et al., in prep.).
Since the data for modeling $\mathrm{C}_2$ excitation came from the ExoMol website based on recent calculations by \citet{2020MNRAS.497.1081M}, updating the values from \citet{2018MNRAS.480.3397Y}, we compared these results with the relevant values used in fitting the spectra to check for consistency.
The wavelengths are consistent, considering the spectral resolution of HPF, and the line oscillator strengths agree at the $1$ to $2\%$ level.

However, we note another point regarding the transition wavenumbers and associated
air wavelengths of Q lines displayed in the ExoMol database, where the discrepancies
between experimental \citep{1977JMoSp..68..399C} and reported values in ExoMol increase with the rotational quantum number.
This is likely due to the presence of $\Lambda$-doubling, which was not taken into account in the ExoMol calculations (Tennyson, Yurchenko, priv. comm.).

The data for $\mathrm{C}_2$ toward HD~24534 are taken from Table\,1 in \citet{2024ApJ...973..143N} and correspond to the average value derived from observations in the ultraviolet F-X and D-X systems with the Space Telescope Imaging Spectrograph
(STIS) and visible Phillips system with the Astrophysical Research Consortium (ARC) echelle spectrograph on the $3.5\,\mathrm{m}$ telescope at the Apache Point Observatory.

\section{Comparison with observations}\label{Comparison}

\subsection{Inversion approach}

Using the various processes presented in Sect.\,\ref{Excitation}, it is possible to solve the detailed balance equations of $\mathrm{C}_2$ for given (fixed) physical conditions. The free parameters are the temperature, $T$, of the gas, its number density, $n_\mathrm{H}$, and a scaling factor, $G$, for the radiation field intensity. If chemical formation excitation is included, then we must also specify the relevant formation rate, $k_f$.

To solve the "inverse problem", we define a ($\chi^2$ like) cost function adapted to excitation diagrams:
\begin{equation}
C \left(T, n_\mathrm{H}, G, k_f \right) = \frac{1}{n}\,\sum_{i=1}^{n}
\left( \frac{
\ln \left( \frac{x_{i}^{obs}}{g_i} \right)
- \ln \left( \frac{x_{i}^{calc}\left(T, n_\mathrm{H}, G, k_f \right)}{g_i} \right)
}{\tau_{i}} \right)^{2},
\end{equation}
where $x_{i}^{obs}$ are estimates of the observed fractional populations derived by normalizing\footnote{This normalization is used for observations in all excitation diagrams also.} the total observed column density to $1$.
Here, $x_{i}^{calc}$ are the fractional populations calculated from the model.
For logarithmic variables, the standard maximum likelihood error is $\tau_i = \sigma_i / x_{i}^{obs}$, where $\sigma_i$ is the observational error. However, observational fitting procedures give uncertainties that are systematically lower for weak lines than for strong lines. To compensate for this bias, we use $\tau_i = \frac{g_i\,\sigma_i}{x_{i}^{obs}}$ in difficult cases.
The number of observed levels is $n$. It is possible to fix one or several parameters to a prescribed value, and optimize the remaining ones.
In Table\,\ref{tab:Derived-physical-parameters.}, we report the derived parameters for the three lines of sight that we discuss below and compare with the values derived in \cite{2024ApJ...973..143N}.

\begin{table}

\caption{Derived physical parameters. \protect\label{tab:Derived-physical-parameters.}}
\centering
\begin{tabular}{ccccc}
\hline\hline
LoS & $T\,\left(\mathrm{K}\right)$ & $n_{\mathrm{H}}\,\left(\mathrm{cm}^{-3}\right)$ & $G$ & $k_{f}\,\left(\mathrm{s}^{-1}\right)$\tabularnewline
\hline
Cyg OB2-12 & $34.7$ & $24.1$ & $0.27$ & $5.6\,10^{-12}$\tabularnewline
        &  40$^{+4}_{-3}$ & 82$^{+2}_{-1}$   &  1.0   &  0    \tabularnewline
HD 29647 & $12.0$ & $70.7$ & $0.48$ & $6\,10^{-12}$\tabularnewline
        &  10$^{+2}_{-1}$ & 122$^{+8}_{-6}$   &  1.0   &  0    \tabularnewline
HD 24534 & $38.0$ & $42.0$ & $0.7$ & -\tabularnewline
        &  46$^{+9}_{-7}$ & 70$^{+4}_{-6}$   &  1.0   &    0  \tabularnewline
\hline
\end{tabular}
\tablefoot{The first line is the present derivation and the second line shows the values deduced in \cite{2024ApJ...973..143N}. Note that $n_{\mathrm{H}}$ and $G$ are degenerate for HD 29647 and HD 24534 (see text)}
\end{table}

\subsection{Cyg OB2-12}\label{Sect:Cyg-OB2}

This line of sight exhibits measurable absorption lines up to $J=32$,
a very significant improvement over the previously reported data by \cite{2024ApJ...973..143N} obtained with ARC echelle data
where the highest $J$ is equal to $14$, and the data from \citet{2019ApJ...881..143H} where the highest level is $J = 28$. However, the $J=30$ level has not been detected due to a blend of Q(30) with a diffuse interstellar bands (DIB) feature.

\begin{figure}
\centering\includegraphics[width=1\columnwidth]{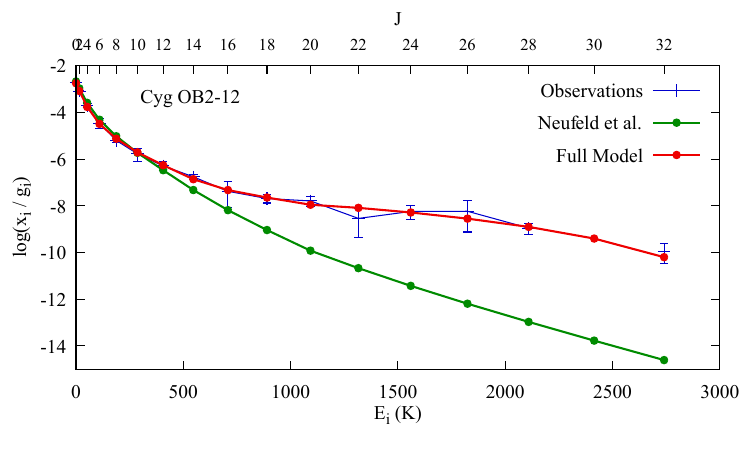}

\caption{
Observed column densities are scaled to $[0:1]$ for comparison with local relative populations $x_J$.
The figure shows a comparison between the present observations and model results for Cyg OB2 12. The observed values are in blue. Model results obtained with the physical parameters derived by \cite{2024ApJ...973..143N} and the present study (displayed in Table\,\ref{tab:Derived-physical-parameters.}) are reported in green and red, respectively. }
\label{Fig:CygOB2}
\end{figure}

In Fig.\,\ref{Fig:CygOB2}, we display the excitation diagram obtained from the present observations and the 0D model results derived from the present fit and those reported in \citet{2024ApJ...973..143N}. We derive a much smaller
density, associated with a smaller radiation field scaling.
The two models display very similar results up to $J = 14$ despite these significant differences.
We do not compare our results with those derived by \cite{2019ApJ...881..143H}
since their diagnostic was obtained with previous collisional rates.

Standard $\chi^2$ theory does not allow derivations of accurate error bars here. However, we find that the temperature and chemical rate are well defined, with typical uncertainties of $\pm 2\,\mathrm{K}$ and $\pm2\,10^{-12}\,\mathrm{s}^{-1}$. The degeneracy between $n_\mathrm{H}$ and $G$ is much harder to lift.
Fig.\,\ref{Fig:Cost_n_G} shows the cost function computed with the values of $T$ and $k_f$ restricted to their best value. We observe a deep valley, which is not straight, but where $n_\mathrm{H}/G \simeq 89$, and which remains shallow over a large range of densities. The best value reported here could easily shift along that valley under the influence of other physical constraints.
We note that using $G=1$ would lead to a value of $\sim 89 \,\mathrm{cm}^{-3}$
for the density, which is close to the value $82\, \mathrm{cm}^{-3}$ derived by
\citet{2024ApJ...973..143N}.
Here we face the limit of such a 0D model that assumes
constant density and temperature.
Using a more sophisticated PDR model is probably the best way to incorporate all the necessary ingredients.

\begin{figure}
\centering\includegraphics[width=1\columnwidth]{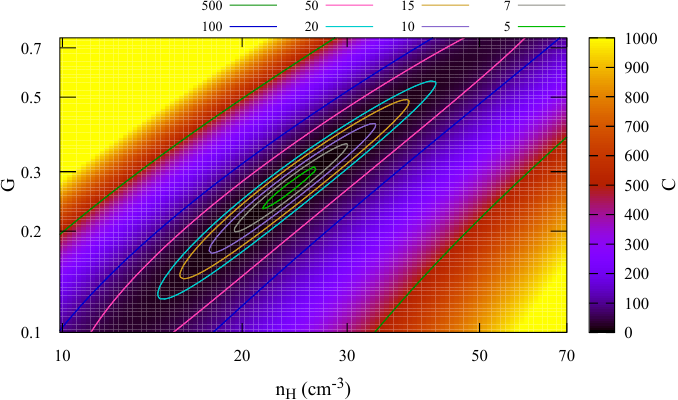}

\caption{Cost function $C$ for Cyg OB2-12 for $T = 34.7\,\mathrm{K}$ and $k_f = 5.6\,10^{-12}\,\mathrm{s}^{-1}$.}
\label{Fig:Cost_n_G}
\end{figure}

\subsection{HD 29647}\label{HD29647}

This line of sight was also reported by \cite{2024ApJ...973..143N}, where values up to $J$ = 12 are available and considered. Our present observations include levels up to J = 24 but the levels $J=20$ and $J=22$ are missing, as shown on Fig.\,\ref{Fig:Adjust_29647_k}.
Direct minimization of a cost function proved difficult, being very sensitive to computation choices and prone to various degeneracies. So, we fixed parameter $k_f$ to a set of values (from $8\,10^{-13}$ to $1.5\,10^{-11}\,\mathrm{s}^{-1}$) and for each $k_f$ we derived the optimal set of $\left( T, n_\mathrm{H}, G \right)$. The results are shown on Fig.\,\ref{Fig:3D_T_n_G-crop}. For each $k_f$, we find a well-defined optimum. However, the cost function variations are well below the uncertainties and do not allow us to pick a specific value. We find that $T$ falls within a very narrow range around $T=12\,\mathrm{K}$, and that, also here, there is a well-defined relation between $n_\mathrm{H}$ and $G$, with $n_\mathrm{H} \simeq 145 \times G\, (\mathrm{cm}^{-3})$.

The set of (degenerate) fits are compared to our observations in Fig.\,\ref{Fig:Adjust_29647_k}. Clearly, an independent argument is needed to lift the degeneracy. Nevertheless, the "best fit" with a negligible chemical excitation is clearly not acceptable. The values shown in Table\,\ref{tab:Derived-physical-parameters.} correspond to the (rather arbitrary) choice $k_f = 6\,10^{-12}\,\mathrm{s}^{-1}$.

\begin{figure}
\centering\includegraphics[width=1\columnwidth]{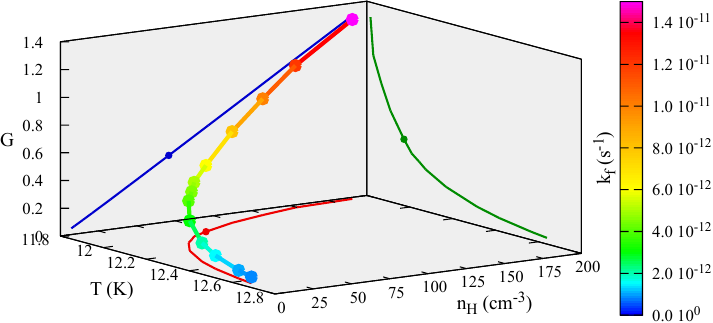}
\caption{
HD 29647 best fit to $\left( T, n_\mathrm{H}, G \right)$ for a range of $k_f$. Values of $k_f$ are color coded. The value selected in Table\,\ref{tab:Derived-physical-parameters.} is shown by a point on the projected curves.
}
\label{Fig:3D_T_n_G-crop}
\end{figure}

\begin{figure}
\centering\includegraphics[width=1\columnwidth]{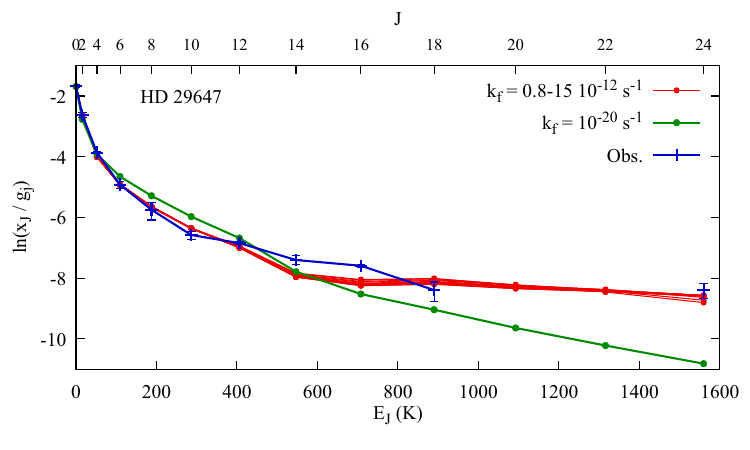}
\caption{
Excitation diagram for $13$ values of $k_f$ going from $0.8\,10^{-12}$ to $15\,10^{-12}\,\mathrm{s}^{-1}$, using optimal parameters from Fig.\,\ref{Fig:3D_T_n_G-crop} (red). The green line corresponds to a negligible chemical excitation.
}
\label{Fig:Adjust_29647_k}
\end{figure}

\subsection{HD 24534}\label{HD24534}

This is the fiducial case studied by \citet{2024ApJ...973..143N} as reported by \citet{2007ApJS..168...58S}.
Since observational column densities are only available up to $J = 16$, we set $k_f = 0$.
This implies that the degeneracy between $n_\mathrm{H}$ and $G$ cannot be lifted, as seen in Fig.\,\ref{Fig:G_nH_9}. The scaling here is $n_\mathrm{H} \simeq 60 \times G \,\mathrm{cm}^{-3}$. Fig.\,\ref{Fig:Adjust_HD24534_4} shows the proposed fit. One can see that levels $J = 0$ and $J = 16$ seem to deviate significantly from the trend of the other levels.

The values reported in Table\,\ref{tab:Derived-physical-parameters.} are only indicative. The associated cloud has an $E_\mathrm{B-V} \simeq 0.62$ \citep{1998ApJ...496L.113S} and is located in the Perseus cloud. We note that $\mathrm{C}_2$ is exposed to a varying radiation field along the line of sight.

\begin{figure}
\centering\includegraphics[width=1\columnwidth]{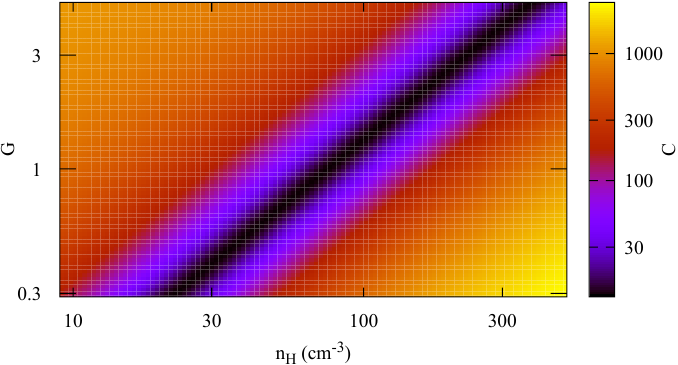}
\caption{Cost function $C$ for $T = 38\,\mathrm{K}$ and $k_f = 0$ for HD 24534. For all values of $n_\mathrm{H} / G \sim 60$, the populations are indistinguishable.}
\label{Fig:G_nH_9}
\end{figure}

\begin{figure}
\centering\includegraphics[width=1\columnwidth]{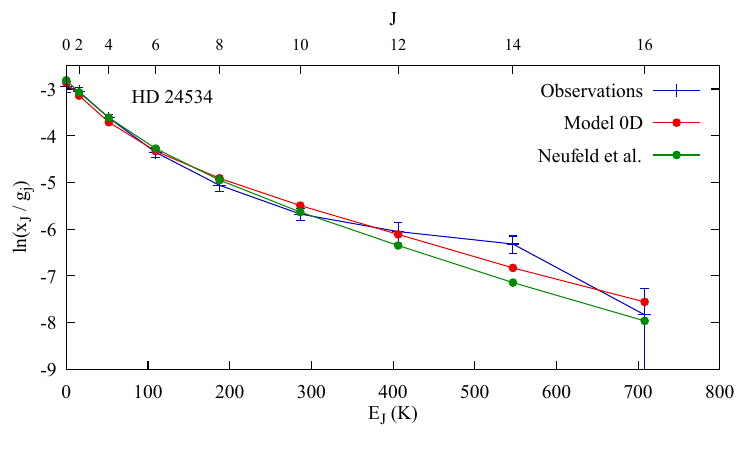}
\caption{HD 24534. Observed column densities are scaled to $[0:1]$ for comparison with local relative populations $x_J$.
Model 0D corresponds to the parameters derived in the present study and points in green are obtained with the parameters derived by \citet{2024ApJ...973..143N} as shown in Table\,\ref{tab:Derived-physical-parameters.}}
\label{Fig:Adjust_HD24534_4}
\end{figure}

\section{Summary}
\label{Discussion}
In this work we present a detailed model of the excitation of interstellar $\mathrm{C}_2$. Compared to previous work, we extend available molecular data using controlled extrapolations to account for the highest rotational levels of $X$ that are now seen in observations. This includes:
\begin{itemize}
    \item Extrapolating and deriving rotational electric quadrupole Einstein emission coefficients within the ground electronic state up to $J = 34$.
    \item Deriving vibrational $1-0$ electric quadrupole Einstein emission
    coefficients up to $v= 1$, $J = 14$.
    \item Extrapolating and deriving collisional de-excitation rate coefficients with $\mathrm{H}_2$ and $\mathrm{He}$ up to $J = 34$ for $\Delta J = 2$, $4$, $6$, and $8$.
\end{itemize}
We have accounted for the electronic radiative transitions between the first eight electronic states of $\mathrm{C}_2$ by using the latest data
available in the ExoMol database. These include in particular the $A-X$ Phillips and the $a-X$ intercombination systems.
We have also introduced the pumping through the $D-X$ Mulliken system in the UV spectral range by using the data reported in \cite{2007JChPh.127w4310S} and \cite{2018JMoSp.344....1K}. We find that their impact is usually negligible.

This all together leads to solving a system of $120$ coupled equations including $X$ rotational levels up to $J=34$ and levels belonging to $a$ that are located between the rotationally excited levels belonging to $X$.
The contribution of radiative pumping and subsequent radiative decay involving the other levels is computed with the radiative cascade formalism. Our results, limited to these processes, are qualitatively similar to those of \citet{2024ApJ...973..143N}.

In addition, our coupled equations include the possibility of direct excitation through exothermic chemical formation processes.
We show that this excitation mechanism allows us to populate high-lying rotational levels up to the observed values, whereas the standard balance between collisional and radiative processes leads to an unavoidable decline of the excitation diagram for high $J$ values. Our prescription -- population proportional to the statistical weight -- makes the minimal amount of assumptions, while remaining plausible. It is clear that
more studies of this specific process are needed to better quantify the effect.

We show that the lowest rotational levels are mainly sensitive to collisions, which constrains the temperature. Intermediate levels are sensitive to the ratio $n_\mathrm{H}/G$, but not to either of them individually. The highest levels are sensitive to chemical excitation. Their new detection
shows that this additional excitation mechanism is relevant for
understanding $\mathrm{C}_2$ excitation.
If enough $J$ levels are observed, then this allows us to determine a unique set of physical parameters, within the approximation of a 0D model. However, if only the lowest levels are seen, typically up to $J=16$, then the degeneracy between density and radiation cannot be lifted.
Comparison with previous studies is thus delicate, as most observations
are restricted up to $J=16$ or less.
The availability of determining the population of
high $J$ levels then reveals new possibilities.
The present trend of our results points to low values of the density,
as already found by \cite{2024ApJ...973..143N} and resulting from
significantly different collisional excitation rate coefficients of $\mathrm{C}_2$ by para and ortho-$\mathrm{H}_2$, compared to previous estimates.
We also point out the sensitivity of our results on the kinetic temperature
that determines the ortho-to-para ratio of $\mathrm{H}_2$, which further impacts the effect of the collisional excitation of $\mathrm{C}_2$.

This study is a significant step toward a better understanding of $\mathrm{C}_2$ excitation but does not account for the variation of density and temperature in the line of sight driven by the absorption of the radiation field along the cloud. Deriving definitive conclusions on the actual value of the density and cosmic ionization rate may then be premature. Introducing the present excitation model into a full PDR model is our next goal and will allow us to include the constraints from other molecular species as well.
Finally, our 0D model also gives the rotational populations in the $a$ state. For the lines of sight that we examined here, their populations are too low to allow for detection.

\begin{acknowledgements}
      This work was supported by the Thematic Action “Physique et Chimie du Milieu Interstellaire” (PCMI) of INSU Programme National “Astro”, with contributions from CNRS Physique \& CNRS Chimie, CEA, and CNES.

      We acknowledge the Texas Advanced Computing Center (TACC) at The University of Texas at Austin for providing high performance computing, visualization, and storage resources that have contributed to the results reported within this paper. These results are based on observations obtained with the Habitable-zone Planet Finder Spectrograph on the HET. The HPF team acknowledges support from NSF grants AST-1006676, AST-1126413, AST-1310885, AST-1517592, AST-1310875, ATI 2009889, ATI-2009982, AST-2108512, and the NASA Astrobiology Institute (NNA09DA76A) in the pursuit of precision radial velocities in the NIR. The HPF team also acknowledges support from the Heising-Simons Foundation via grant 2017-0494.
\end{acknowledgements}

\bibliographystyle{aa}
\bibliography{aa54760-25}

\begin{appendix}

\section{$\mathrm{C}_2$ data}\label{Sec:C2_data}

\subsection{Energy levels}\label{Energies}

Energies, in $\mathrm{cm}^{-1}$, are in Table\,\ref{tab:Energy_Levels}.

\begin{table}[h!]
\caption{$\mathrm{C}_{2}$ rotational level energies 
\protect\label{tab:Energy_Levels}}

\centering
\begin{tabular}{cc|cc}
\hline \hline
$v=0,\,J$ & $E_{J}$ & $J$ & $E_{J}$\tabularnewline
\hline
$0$ & $0$ & $18$ & $618.574842$\tabularnewline
$2$ & $10.866355$ & $20$ & $759.423253$\tabularnewline
$4$ & $36.219214$ & $22$ & $914.618665$\tabularnewline
$6$ & $76.053856$ & $24$ & $1084.132143$\tabularnewline
$8$ & $130.362858$ & $26$ & $1267.932072$\tabularnewline
$10$ & $199.136103$ & $28$ & $1465.984169$\tabularnewline
$12$ & $282.360777$ & $30$ & $1678.251483$\tabularnewline
$14$ & $380.021373$ & $32$ & $1904.694400$\tabularnewline
$16$ & $492.099690$ & $34$ & $2145.270648$\tabularnewline
\hline
\end{tabular}

\begin{tabular}{cc}
\hline
$v=1,\,J$ & $E_{J}$\tabularnewline
\hline
$0$ & $1827.487599$\tabularnewline
$2$ & $1838.238745$\tabularnewline
$4$ & $1863.337620$\tabularnewline
$6$ & $1902.771206$\tabularnewline
$8$ & $1956.533032$\tabularnewline
$10$ & $2024.615469$\tabularnewline
$12$ & $2107.001692$\tabularnewline
$14$ & $2203.678462$\tabularnewline
 & \tabularnewline
\hline
\end{tabular}
\tablefoot{Rotational energies (in $\mathrm{cm}^{-1}$) of $\mathrm{C}_{2}$
first two vibrational levels, up to $E=2204\, \mathrm{cm}^{-1}$, from \href{https://www.ExoMol.com/data/molecules/C2/12C2/8states/}{https://www.ExoMol.com/}}
\end{table}

\subsection{Electric quadrupole Einstein coefficients}\label{Quad_Coeff}

Einstein emission coefficients are in $\mathrm{s}^{-1}$, and corresponding wavenumbers in $\mathrm{cm}^{-1}$. See Table\,\ref{tab:Quadrupolar-transitions-proba} and \ref{tab:v_1-0_Aul}.

\begin{table}[h!]

\caption{$\mathrm{C}_{2}$ electric quadrupole emission coefficients \protect\label{tab:Quadrupolar-transitions-proba}}

\centering
\begin{tabular}{cccc}
\hline
\hline
$J_{u}$ & $J_{l}$ & $A_{ul} \, \left(\mathrm{s}^{-1}\right)$ & $\bar{\nu}_{ul} \,\left(\mathrm{cm}^{-1}\right)$\tabularnewline
\hline
$2$ & $0$ & $1.763\,10^{-17}$* & $10.875$\tabularnewline
$4$ & $2$ & $1.742\,10^{-15}$* & $25.338$\tabularnewline
$6$ & $4$ & $1.838\,10^{-14}$* & $39.836$\tabularnewline
$8$ & $6$ & $9.065\,10^{-14}$* & $54.316$\tabularnewline
$10$ & $8$ & $3.034\,10^{-13}$* & $68.766$\tabularnewline
$12$ & $10$ & $8.020\,10^{-13}$* & $83.223$\tabularnewline
$14$ & $12$ & $1.808\,10^{-12}$* & $97.658$\tabularnewline
$16$ & $14$ & $3.638\,10^{-12}$* & $112.090$\tabularnewline
$18$ & $16$ & $6.712\,10^{-12}$* & $126.467$\tabularnewline
$20$ & $18$ & $1.158\,10^{-11}$* & $140.846$\tabularnewline
$22$ & $20$ & $1.914\,10^{-11}$ & $155.200$\tabularnewline
$24$ & $22$ & $3.016\,10^{-11}$ & $169.510$\tabularnewline
$26$ & $24$ & $4.582\,10^{-11}$ & $183.797$\tabularnewline
$28$ & $26$ & $6.750\,10^{-11}$ & $198.054$\tabularnewline
$30$ & $28$ & $9.681\,10^{-11}$ & $212.265$\tabularnewline
$32$ & $30$ & $1.356\,10^{-10}$ & $226.440$\tabularnewline
$34$ & $32$ & $1.862\,10^{-10}$ & $240.575$\tabularnewline
\hline
\end{tabular}
\tablefoot{Extrapolated (present work) rotational electric quadrupole
Einstein emission coefficients within $v=0$ in $\mathrm{s}^{-1}$
and corresponding wavenumbers (in $\mathrm{cm}^{-1}$). Values with a star are from \citet{1982ApJ...258..533V}.}
\end{table}

\begin{table}[h!]

\caption{$v=1\rightarrow0$ electric quadrupole emission coefficients
\protect\label{tab:v_1-0_Aul}
}

\centering
\begin{tabular}{ccccccc}
\hline\hline
$J'$ & \multicolumn{2}{c}{$J"=J'-2$} & \multicolumn{2}{c}{$J"=J'$} & \multicolumn{2}{c}{$J"=J+2$}\tabularnewline
 & $A_{ul}$ & $\bar{\nu}_{ul}$ & $A_{ul}$ & $\bar{\nu}_{ul}$ & $A_{ul}$ & $\bar{\nu}_{ul}$\tabularnewline
\hline
$0$ & - & - & - & - & $6.46$ & $1816.61$\tabularnewline
$2$ & $1.29$ & $1838.24$ & $1.85$ & $1827.36$ & $3.32$ & $1802.03$\tabularnewline
$4$ & $1.85$ & $1852.46$ & $1.68$ & $1827.12$ & $2.94$ & $1787.29$\tabularnewline
$6$ & $2.03$ & $1866.56$ & $1.64$ & $1826.72$ & $2.78$ & $1772.41$\tabularnewline
$8$ & $2.13$ & $1880.48$ & $1.63$ & $1826.17$ & $2.70$ & $1757.40$\tabularnewline
$10$ & $2.19$ & $1894.25$ & $1.63$ & $1825.48$ & $2.65$ & $1742.26$\tabularnewline
$12$ & $2.22$ & $1907.87$ & $1.62$ & $1824.65$ & $2.61$ & $1726.99$\tabularnewline
$14$ & $2.25$ & $1921.32$ & $1.62$ & $1823.67$ & $2.59$ & $1711.58$\tabularnewline
\hline
\end{tabular}
\tablefoot{$v=1\rightarrow0$ electric quadrupole Einstein emission coefficients,
 derived in present work, in $10^{-8}\,\mathrm{s}^{-1}$, and corresponding wavenumbers $\bar{\nu}_{ul}$ in $\mathrm{cm}^{-1}$}
\end{table}

\subsection{Collision rates extrapolations}\label{Sec:Coll_rates}

Collision rate coefficients with $\mathrm{H}_2$, in $\mathrm{cm}^{3}\,\mathrm{s}^{-1}$, are in Table\,\ref{tab:Ortho-H2-Col} and \ref{tab:Para-H2-Col}. Rate coefficients with $\mathrm{He}$ are in Table\,\ref{tab:He-Col}.

\begin{sidewaystable*}

Table\,\ref{tab:Ortho-H2-Col} and \ref{tab:Para-H2-Col} give the rate coefficients shown on Fig.\,\ref{fig:Extrapolation-of-collision}. Numbers in parenthesis are powers of $10$.

\caption{Ortho $\mathrm{H}_{2}$ collisional de-excitation rate coefficients 
\protect\label{tab:Ortho-H2-Col}}
\bigskip{}
\begin{tabular}{cccccccccccc}
\hline\hline
$J'$ & $J"$ & $10$ & $20$ & $30$ & $40$ & $50$ & $60$ & $70$ & $80$ & $90$ & $100$\tabularnewline
\hline
$22$ & $14$ & $3.4428(-17)$ & $6.6970(-16)$ & $1.5765(-15)$ & $1.9819(-15)$ & $1.8688(-15)$ & $1.5158(-15)$ & $1.1351(-15)$ & $8.1915(-16)$ & $5.8389(-16)$ & $4.1731(-16)$\tabularnewline
$22$ & $16$ & $6.8094(-15)$ & $3.0495(-14)$ & $4.4899(-14)$ & $5.0013(-14)$ & $4.9727(-14)$ & $4.6604(-14)$ & $4.2205(-14)$ & $3.7397(-14)$ & $3.2633(-14)$ & $2.8254(-14)$\tabularnewline
$22$ & $18$ & $2.2598(-13)$ & $5.7749(-13)$ & $8.9647(-13)$ & $9.5671(-13)$ & $9.4878(-13)$ & $9.0807(-13)$ & $8.5169(-13)$ & $7.8862(-13)$ & $7.2387(-13)$ & $6.6076(-13)$\tabularnewline
$22$ & $20$ & $5.5006(-12)$ & $9.7658(-12)$ & $1.1213(-11)$ & $1.1801(-11)$ & $1.2075(-11)$ & $1.2205(-11)$ & $1.2237(-11)$ & $1.2203(-11)$ & $1.2122(-11)$ & $1.1984(-11)$\tabularnewline
$24$ & $16$ & $1.6177(-17)$ & $4.7820(-16)$ & $1.2712(-15)$ & $1.6434(-15)$ & $1.5244(-15)$ & $1.1883(-15)$ & $8.4499(-16)$ & $5.7612(-16)$ & $3.8759(-16)$ & $2.6183(-16)$\tabularnewline
$24$ & $18$ & $4.7565(-15)$ & $2.5911(-14)$ & $4.0039(-14)$ & $4.4998(-14)$ & $4.4388(-14)$ & $4.0907(-14)$ & $3.6248(-14)$ & $3.1332(-14)$ & $2.6622(-14)$ & $2.2428(-14)$\tabularnewline
$24$ & $20$ & $1.7182(-13)$ & $4.9549(-13)$ & $8.2210(-13)$ & $8.8025(-13)$ & $8.6582(-13)$ & $8.1731(-13)$ & $7.5374(-13)$ & $6.8505(-13)$ & $6.1658(-13)$ & $5.5164(-13)$\tabularnewline
$24$ & $22$ & $4.1612(-12)$ & $7.8235(-12)$ & $9.0873(-12)$ & $9.5902(-12)$ & $9.8120(-12)$ & $9.9058(-12)$ & $9.9124(-12)$ & $9.8610(-12)$ & $9.7688(-12)$ & $9.6274(-12)$\tabularnewline
$26$ & $18$ & $7.6017(-18)$ & $3.4146(-16)$ & $1.0250(-15)$ & $1.3627(-15)$ & $1.2435(-15)$ & $9.3154(-16)$ & $6.2902(-16)$ & $4.0519(-16)$ & $2.5729(-16)$ & $1.6428(-16)$\tabularnewline
$26$ & $20$ & $3.3225(-15)$ & $2.2016(-14)$ & $3.5705(-14)$ & $4.0486(-14)$ & $3.9623(-14)$ & $3.5907(-14)$ & $3.1131(-14)$ & $2.6250(-14)$ & $2.1718(-14)$ & $1.7803(-14)$\tabularnewline
$26$ & $22$ & $1.3065(-13)$ & $4.2513(-13)$ & $7.5390(-13)$ & $8.0990(-13)$ & $7.9011(-13)$ & $7.3562(-13)$ & $6.6705(-13)$ & $5.9508(-13)$ & $5.2518(-13)$ & $4.6055(-13)$\tabularnewline
$26$ & $24$ & $3.1479(-12)$ & $6.2674(-12)$ & $7.3648(-12)$ & $7.7935(-12)$ & $7.9735(-12)$ & $8.0399(-12)$ & $8.0294(-12)$ & $7.9682(-12)$ & $7.8723(-12)$ & $7.7341(-12)$\tabularnewline
$28$ & $20$ & $3.5720(-18)$ & $2.4382(-16)$ & $8.2654(-16)$ & $1.1300(-15)$ & $1.0144(-15)$ & $7.3028(-16)$ & $4.6825(-16)$ & $2.8497(-16)$ & $1.7080(-16)$ & $1.0307(-16)$\tabularnewline
$28$ & $22$ & $2.3208(-15)$ & $1.8706(-14)$ & $3.1841(-14)$ & $3.6426(-14)$ & $3.5369(-14)$ & $3.1517(-14)$ & $2.6737(-14)$ & $2.1993(-14)$ & $1.7717(-14)$ & $1.4132(-14)$\tabularnewline
$28$ & $24$ & $9.9337(-14)$ & $3.6477(-13)$ & $6.9136(-13)$ & $7.4518(-13)$ & $7.2103(-13)$ & $6.6210(-13)$ & $5.9034(-13)$ & $5.1692(-13)$ & $4.4734(-13)$ & $3.8449(-13)$\tabularnewline
$28$ & $26$ & $2.3813(-12)$ & $5.0209(-12)$ & $5.9688(-12)$ & $6.3335(-12)$ & $6.4795(-12)$ & $6.5256(-12)$ & $6.5041(-12)$ & $6.4387(-12)$ & $6.3440(-12)$ & $6.2132(-12)$\tabularnewline
$30$ & $22$ & $1.6785(-18)$ & $1.7410(-16)$ & $6.6648(-16)$ & $9.3696(-16)$ & $8.2748(-16)$ & $5.7250(-16)$ & $3.4857(-16)$ & $2.0043(-16)$ & $1.1338(-16)$ & $6.4670(-17)$\tabularnewline
$30$ & $24$ & $1.6211(-15)$ & $1.5894(-14)$ & $2.8394(-14)$ & $3.2774(-14)$ & $3.1572(-14)$ & $2.7665(-14)$ & $2.2963(-14)$ & $1.8427(-14)$ & $1.4453(-14)$ & $1.1218(-14)$\tabularnewline
$30$ & $26$ & $7.5531(-14)$ & $3.1297(-13)$ & $6.3401(-13)$ & $6.8562(-13)$ & $6.5798(-13)$ & $5.9592(-13)$ & $5.2244(-13)$ & $4.4903(-13)$ & $3.8103(-13)$ & $3.2099(-13)$\tabularnewline
$30$ & $28$ & $1.8015(-12)$ & $4.0222(-12)$ & $4.8374(-12)$ & $5.1470(-12)$ & $5.2654(-12)$ & $5.2964(-12)$ & $5.2685(-12)$ & $5.2028(-12)$ & $5.1124(-12)$ & $4.9913(-12)$\tabularnewline
$32$ & $24$ & $7.8872e-19$ & $1.2432(-16)$ & $5.3741(-16)$ & $7.7692(-16)$ & $6.7500(-16)$ & $4.4881(-16)$ & $2.5948(-16)$ & $1.4096(-16)$ & $7.5262(-17)$ & $4.0576(-17)$\tabularnewline
$32$ & $26$ & $1.1324(-15)$ & $1.3505(-14)$ & $2.5321(-14)$ & $2.9487(-14)$ & $2.8182(-14)$ & $2.4283(-14)$ & $1.9722(-14)$ & $1.5438(-14)$ & $1.1791(-14)$ & $8.9048(-15)$\tabularnewline
$32$ & $28$ & $5.7431(-14)$ & $2.6853(-13)$ & $5.8142(-13)$ & $6.3083(-13)$ & $6.0045(-13)$ & $5.3636(-13)$ & $4.6236(-13)$ & $3.9006(-13)$ & $3.2455(-13)$ & $2.6798(-13)$\tabularnewline
$32$ & $30$ & $1.3628(-12)$ & $3.2222(-12)$ & $3.9204(-12)$ & $4.1827(-12)$ & $4.2788(-12)$ & $4.2988(-12)$ & $4.2676(-12)$ & $4.2042(-12)$ & $4.1199(-12)$ & $4.0097(-12)$\tabularnewline
$34$ & $26$ & $3.7062e-19$ & $8.8768(-17)$ & $4.3334(-16)$ & $6.4423(-16)$ & $5.5062(-16)$ & $3.5185(-16)$ & $1.9316(-16)$ & $9.9140(-17)$ & $4.9961(-17)$ & $2.5458(-17)$\tabularnewline
$34$ & $28$ & $7.9099(-16)$ & $1.1475(-14)$ & $2.2580(-14)$ & $2.6530(-14)$ & $2.5157(-14)$ & $2.1315(-14)$ & $1.6938(-14)$ & $1.2935(-14)$ & $9.6186(-15)$ & $7.0686(-15)$\tabularnewline
$34$ & $30$ & $4.3668(-14)$ & $2.3040(-13)$ & $5.3319(-13)$ & $5.8041(-13)$ & $5.4795(-13)$ & $4.8275(-13)$ & $4.0918(-13)$ & $3.3883(-13)$ & $2.7645(-13)$ & $2.2373(-13)$\tabularnewline
$34$ & $32$ & $1.0309(-12)$ & $2.5813(-12)$ & $3.1773(-12)$ & $3.3991(-12)$ & $3.4770(-12)$ & $3.4891(-12)$ & $3.4569(-12)$ & $3.3972(-12)$ & $3.3200(-12)$ & $3.2212(-12)$\tabularnewline
\hline
\end{tabular}
\tablefoot{Extrapolation of rotational de-excitation rate coefficients of $\mathrm{C}_2$ due to Ortho $\mathrm{H}_{2}$ in $\mathrm{cm}^3\,\mathrm{s}^{-1}$.
Column headers $3$ to $12$ are temperatures in $\mathrm{K}$.}
\end{sidewaystable*}

\begin{sidewaystable*}
\caption{Para $\mathrm{H}_{2}$ collisional de-excitation rate coefficients 
\protect\label{tab:Para-H2-Col}}
\bigskip{}
\begin{tabular}{cccccccccccc}
\hline\hline
$J'$ & $J"$ & $10$ & $20$ & $30$ & $40$ & $50$ & $60$ & $70$ & $80$ & $90$ & $100$\tabularnewline
\hline
$22$ & $14$ & $8.8245(-18)$ & $1.4483(-16)$ & $3.4664(-16)$ & $5.4216(-16)$ & $7.1989(-16)$ & $8.7989(-16)$ & $1.0188(-15)$ & $1.1281(-15)$ & $1.2003(-15)$ & $1.2367(-15)$\tabularnewline
$22$ & $16$ & $2.3106(-15)$ & $9.6025(-15)$ & $2.2897(-14)$ & $3.4420(-14)$ & $4.3640(-14)$ & $5.0959(-14)$ & $5.6665(-14)$ & $6.0807(-14)$ & $6.3474(-14)$ & $6.4726(-14)$\tabularnewline
$22$ & $18$ & $2.8808(-14)$ & $1.5657(-13)$ & $2.6191(-13)$ & $3.3984(-13)$ & $4.0280(-13)$ & $4.5756(-13)$ & $5.0695(-13)$ & $5.5157(-13)$ & $5.9155(-13)$ & $6.2586(-13)$\tabularnewline
$22$ & $20$ & $2.7787(-12)$ & $5.5406(-12)$ & $6.7424(-12)$ & $7.5528(-12)$ & $8.2617(-12)$ & $8.9591(-12)$ & $9.6609(-12)$ & $1.0371(-11)$ & $1.1070(-11)$ & $1.1761(-11)$\tabularnewline
$24$ & $16$ & $4.6216(-18)$ & $1.0670(-16)$ & $2.8401(-16)$ & $4.6800(-16)$ & $6.4055(-16)$ & $7.9779(-16)$ & $9.3408(-16)$ & $1.0391(-15)$ & $1.1045(-15)$ & $1.1318(-15)$\tabularnewline
$24$ & $18$ & $1.7199(-15)$ & $8.2508(-15)$ & $2.2391(-14)$ & $3.5589(-14)$ & $4.6377(-14)$ & $5.4886(-14)$ & $6.1339(-14)$ & $6.5764(-14)$ & $6.8288(-14)$ & $6.9029(-14)$\tabularnewline
$24$ & $20$ & $1.7070(-14)$ & $1.1915(-13)$ & $2.1497(-13)$ & $2.8908(-13)$ & $3.4980(-13)$ & $4.0273(-13)$ & $4.5030(-13)$ & $4.9294(-13)$ & $5.3076(-13)$ & $5.6255(-13)$\tabularnewline
$24$ & $22$ & $2.3181(-12)$ & $4.9275(-12)$ & $6.0986(-12)$ & $6.9023(-12)$ & $7.6163(-12)$ & $8.3278(-12)$ & $9.0513(-12)$ & $9.7898(-12)$ & $1.0522(-11)$ & $1.1250(-11)$\tabularnewline
$26$ & $18$ & $2.4204(-18)$ & $7.8609(-17)$ & $2.3270(-16)$ & $4.0399(-16)$ & $5.6995(-16)$ & $7.2335(-16)$ & $8.5640(-16)$ & $9.5709(-16)$ & $1.0163(-15)$ & $1.0358(-15)$\tabularnewline
$26$ & $20$ & $1.2803(-15)$ & $7.0894(-15)$ & $2.1896(-14)$ & $3.6797(-14)$ & $4.9284(-14)$ & $5.9115(-14)$ & $6.6398(-14)$ & $7.1125(-14)$ & $7.3467(-14)$ & $7.3618(-14)$\tabularnewline
$26$ & $22$ & $1.0115(-14)$ & $9.0682(-14)$ & $1.7645(-13)$ & $2.4589(-13)$ & $3.0377(-13)$ & $3.5447(-13)$ & $3.9999(-13)$ & $4.4054(-13)$ & $4.7622(-13)$ & $5.0564(-13)$\tabularnewline
$26$ & $24$ & $1.9339(-12)$ & $4.3822(-12)$ & $5.5163(-12)$ & $6.3079(-12)$ & $7.0212(-12)$ & $7.7411(-12)$ & $8.4802(-12)$ & $9.2412(-12)$ & $1.0001(-11)$ & $1.0762(-11)$\tabularnewline
$28$ & $20$ & $1.2676(-18)$ & $5.7914(-17)$ & $1.9066(-16)$ & $3.4873(-16)$ & $5.0713(-16)$ & $6.5585(-16)$ & $7.8518(-16)$ & $8.8157(-16)$ & $9.3519(-16)$ & $9.4789(-16)$\tabularnewline
$28$ & $22$ & $9.5300(-16)$ & $6.0914(-15)$ & $2.1412(-14)$ & $3.8047(-14)$ & $5.2374(-14)$ & $6.3670(-14)$ & $7.1875(-14)$ & $7.6923(-14)$ & $7.9039(-14)$ & $7.8513(-14)$\tabularnewline
$28$ & $24$ & $5.9934(-15)$ & $6.9014(-14)$ & $1.4483(-13)$ & $2.0916(-13)$ & $2.6380(-13)$ & $3.1200(-13)$ & $3.5529(-13)$ & $3.9372(-13)$ & $4.2728(-13)$ & $4.5449(-13)$\tabularnewline
$28$ & $26$ & $1.6134(-12)$ & $3.8973(-12)$ & $4.9895(-12)$ & $5.7647(-12)$ & $6.4727(-12)$ & $7.1956(-12)$ & $7.9452(-12)$ & $8.7234(-12)$ & $9.5062(-12)$ & $1.0294(-11)$\tabularnewline
$30$ & $22$ & $6.6390(-19)$ & $4.2667(-17)$ & $1.5621(-16)$ & $3.0103(-16)$ & $4.5124(-16)$ & $5.9465(-16)$ & $7.1988(-16)$ & $8.1202(-16)$ & $8.6054(-16)$ & $8.6747(-16)$\tabularnewline
$30$ & $24$ & $7.0938(-16)$ & $5.2340(-15)$ & $2.0939(-14)$ & $3.9339(-14)$ & $5.5658(-14)$ & $6.8576(-14)$ & $7.7803(-14)$ & $8.3194(-14)$ & $8.5033(-14)$ & $8.3733(-14)$\tabularnewline
$30$ & $26$ & $3.5514(-15)$ & $5.2523(-14)$ & $1.1888(-13)$ & $1.7792(-13)$ & $2.2909(-13)$ & $2.7461(-13)$ & $3.1559(-13)$ & $3.5187(-13)$ & $3.8337(-13)$ & $4.0851(-13)$\tabularnewline
$30$ & $28$ & $1.3460(-12)$ & $3.4660(-12)$ & $4.5131(-12)$ & $5.2683(-12)$ & $5.9670(-12)$ & $6.6886(-12)$ & $7.4439(-12)$ & $8.2345(-12)$ & $9.0356(-12)$ & $9.8476(-12)$\tabularnewline
$32$ & $24$ & $3.4770(-19)$ & $3.1434(-17)$ & $1.2799(-16)$ & $2.5986(-16)$ & $4.0150(-16)$ & $5.3917(-16)$ & $6.6001(-16)$ & $7.4795(-16)$ & $7.9185(-16)$ & $7.9388(-16)$\tabularnewline
$32$ & $26$ & $5.2805(-16)$ & $4.4972(-15)$ & $2.0476(-14)$ & $4.0675(-14)$ & $5.9147(-14)$ & $7.3860(-14)$ & $8.4220(-14)$ & $8.9976(-14)$ & $9.1482(-14)$ & $8.9300(-14)$\tabularnewline
$32$ & $28$ & $2.1044(-15)$ & $3.9972(-14)$ & $9.7574(-14)$ & $1.5134(-13)$ & $1.9895(-13)$ & $2.4170(-13)$ & $2.8033(-13)$ & $3.1447(-13)$ & $3.4397(-13)$ & $3.6719(-13)$\tabularnewline
$32$ & $30$ & $1.1229(-12)$ & $3.0825(-12)$ & $4.0821(-12)$ & $4.8146(-12)$ & $5.5008(-12)$ & $6.2173(-12)$ & $6.9742(-12)$ & $7.7731(-12)$ & $8.5883(-12)$ & $9.4201(-12)$\tabularnewline
$34$ & $26$ & $1.8210(-19)$ & $2.3159(-17)$ & $1.0487(-16)$ & $2.2431(-16)$ & $3.5725(-16)$ & $4.8886(-16)$ & $6.0512(-16)$ & $6.8894(-16)$ & $7.2864(-16)$ & $7.2652(-16)$\tabularnewline
$34$ & $28$ & $3.9306(-16)$ & $3.8642(-15)$ & $2.0024(-14)$ & $4.2056(-14)$ & $6.2856(-14)$ & $7.9552(-14)$ & $9.1166(-14)$ & $9.7310(-14)$ & $9.8420(-14)$ & $9.5237(-14)$\tabularnewline
$34$ & $30$ & $1.2469(-15)$ & $3.0421(-14)$ & $8.0088(-14)$ & $1.2873(-13)$ & $1.7277(-13)$ & $2.1274(-13)$ & $2.4901(-13)$ & $2.8104(-13)$ & $3.0863(-13)$ & $3.3004(-13)$\tabularnewline
$34$ & $32$ & $9.3679(-13)$ & $2.7413(-12)$ & $3.6923(-12)$ & $4.4000(-12)$ & $5.0710(-12)$ & $5.7793(-12)$ & $6.5342(-12)$ & $7.3375(-12)$ & $8.1632(-12)$ & $9.0112(-12)$\tabularnewline
\hline
\end{tabular}
\tablefoot{Extrapolation of rotational de-excitation rate coefficients of $\mathrm{C}_2$ due to Para $\mathrm{H}_{2}$ in $\mathrm{cm}^3\,\mathrm{s}^{-1}$.
Column headers $3$ to $12$ are temperatures in $\mathrm{K}$.}
\end{sidewaystable*}

\begin{table*}[h!]
\caption{$\mathrm{He}$ collisional de-excitation rate coefficients 
\protect\label{tab:He-Col}}
\centering
\begin{tabular}{cccccccc}
\hline\hline
$J'$ & $J"$ & $5$ & $20$ & $40$ & $60$ & $80$ & $100$\tabularnewline
\hline
$22$ & $18$ & $7.997(-13)$ & $9.994(-13)$ & $1.545(-12)$ & $2.238(-12)$ & $3.051(-12)$ & $3.954(-12)$\tabularnewline
$22$ & $20$ & $7.176(-12)$ & $8.970(-12)$ & $1.409(-11)$ & $1.937(-11)$ & $2.482(-11)$ & $3.024(-11)$\tabularnewline
$24$ & $20$ & $7.236(-13)$ & $9.043(-13)$ & $1.398(-12)$ & $2.025(-12)$ & $2.761(-12)$ & $3.577(-12)$\tabularnewline
$24$ & $22$ & $6.131(-12)$ & $7.663(-12)$ & $1.256(-11)$ & $1.763(-11)$ & $2.294(-11)$ & $2.829(-11)$\tabularnewline
$26$ & $22$ & $6.547(-13)$ & $8.183(-13)$ & $1.265(-12)$ & $1.832(-12)$ & $2.498(-12)$ & $3.237(-12)$\tabularnewline
$26$ & $24$ & $5.237(-12)$ & $6.547(-12)$ & $1.119(-11)$ & $1.605(-11)$ & $2.121(-11)$ & $2.646(-11)$\tabularnewline
$28$ & $24$ & $5.924(-13)$ & $7.404(-13)$ & $1.145(-12)$ & $1.658(-12)$ & $2.260(-12)$ & $2.929(-12)$\tabularnewline
$28$ & $26$ & $4.474(-12)$ & $5.593(-12)$ & $9.968(-12)$ & $1.460(-11)$ & $1.961(-11)$ & $2.475(-11)$\tabularnewline
$30$ & $26$ & $5.360(-13)$ & $6.699(-13)$ & $1.036(-12)$ & $1.500(-12)$ & $2.045(-12)$ & $2.650(-12)$\tabularnewline
$30$ & $28$ & $3.822(-12)$ & $4.778(-12)$ & $8.881(-12)$ & $1.329(-11)$ & $1.812(-11)$ & $2.315(-11)$\tabularnewline
$32$ & $28$ & $4.850(-13)$ & $6.062(-13)$ & $9.371(-13)$ & $1.358(-12)$ & $1.851(-12)$ & $2.398(-12)$\tabularnewline
$32$ & $30$ & $3.265(-12)$ & $4.082(-12)$ & $7.913(-12)$ & $1.210(-11)$ & $1.675(-11)$ & $2.166(-11)$\tabularnewline
$34$ & $30$ & $4.389(-13)$ & $5.485(-13)$ & $8.480(-13)$ & $1.228(-12)$ & $1.675(-12)$ & $2.170(-12)$\tabularnewline
$34$ & $32$ & $2.790(-12)$ & $3.487(-12)$ & $7.050(-12)$ & $1.101(-11)$ & $1.549(-11)$ & $2.026(-11)$\tabularnewline
\hline
\end{tabular}
\tablefoot{Extrapolated $\mathrm{He}$ collision rate coefficients in $\mathrm{cm}^{3}\,\mathrm{s}^{-1}$.
Column headers $3$ to $8$ are temperatures in $\mathrm{K}$.}
\end{table*}

\section{Cascade coefficients}\label{Cascade_Coef}

A cascade coefficient $\mathcal{C}_{nj}$
is the probability that a species on level $n>M$ ends on level $j\leqslant M$
after a cascade through any possible path, where $M$ is the number of levels that are involved in the balance equations \ref{Eq:Steady_state}. That probability can be
computed using the usual properties of probabilities: it is the sum
over all possible intermediate levels of going from $n$ to some level
$m>j$ by any path, followed by a direct single transition from $m$
to $j$. That is:
\begin{equation}
\mathcal{C}_{nj}=\sum_{m=M+1}^{n}\mathcal{A}_{n,m}\,\frac{A_{m,j}}{A_{m}}=\sum_{m=M+1}^{n}\mathcal{A}_{n,m}\,P_{m,j}\label{eq:_final_C}
\end{equation}
where $A_{m,j}$ is the standard Einstein coefficient and $A_{m}$
is the inverse radiative lifetime of level $m$, and we have written
$P_{m,j}={\displaystyle \frac{A_{m,j}}{A_{m}}}$. $\mathcal{A}_{n,m}$
is the probability to go from $n$ to $m$ by any path and we have
$\mathcal{A}_{n,n}=1$. These coefficients follow a very similar relation:
\begin{equation}
\mathcal{A}_{n,m}=\sum_{l=m+1}^{n}\mathcal{A}_{n,l}\,\frac{A_{l,m}}{A_{l}}=\sum_{l=m+1}^{n}\mathcal{A}_{n,l}\,P_{l,m}\label{eq:_recur}
\end{equation}

The only difference is that summation starts at $M+1$ for $\mathcal{C}_{nj}$
and starts at $m+1$ for $\mathcal{A}_{n,m}$. These coefficients
may be computed by recurrence. From any $n>M$ starting from $N$:
\begin{itemize}
\item Set $\mathcal{A}_{n,n}=1$
\item Then $\mathcal{A}_{n,n-1}=\mathcal{A}_{n,n}\,{\displaystyle P_{n,n-1}}$,
$\mathcal{A}_{n,n-2}=\mathcal{A}_{n,n}\,{\displaystyle P_{n,n-1}}+\mathcal{A}_{n,n-1}\,{\displaystyle P_{n,n-2}}$
and apply Eq~\ref{eq:_recur} up to $m=M+1$
\item Once all pairs $\left(n,m\right)$ are known, use Eq~\ref{eq:_final_C}
for the final coefficients $\mathcal{C}_{nj}$.
\end{itemize}

\section{Parameters variations at $k_f = 0$}\label{Variations}

To illustrate that only chemical excitation is efficient enough to reproduce the observed population of high $J$ levels, we show here the impact of varying the $3$ other parameters ($n_\mathrm{H}$, $T$ and $G$) around a base model. We use as a reference $T = 40\,\mathrm{K}$, $n_\mathrm{H} = 100\,\mathrm{cm}^{-3}$ and $G = 1$, and we vary each parameter by about $20\%$ around these values. Figs.\,\ref{Fig:Variations_T}, \ref{Fig:Variations_n} and \ref{Fig:Variations_G} show the impact. It is clear that no choice of these parameters is able to give an almost constant value of $N_J/g_J$ for high $J$, as observed. Chemical excitation succeeds because the state specific formation rate is assumed to be proportional to the statistical weight $g_J = 2J+1$. As deexcitation is roughly the same for all high lying levels, this results in a similar value for $N_J/g_J$.

\begin{figure}[h!]
\centering\includegraphics[width=1\columnwidth]{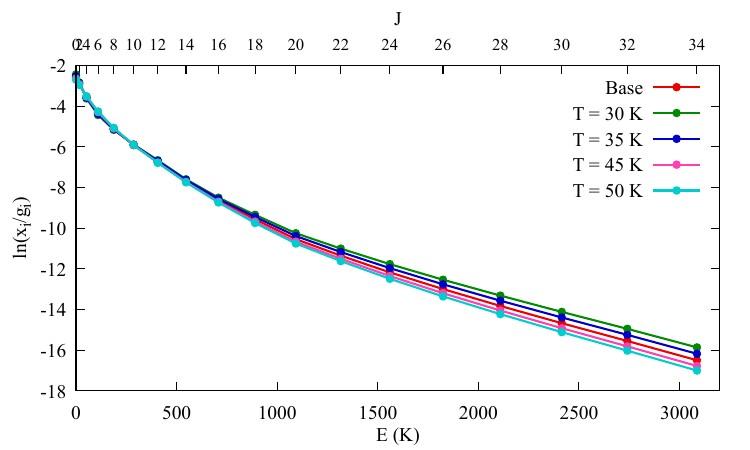}
\caption{Variation of $T$
}
\label{Fig:Variations_T}
\end{figure}
\begin{figure}[h!]
\centering\includegraphics[width=1\columnwidth]{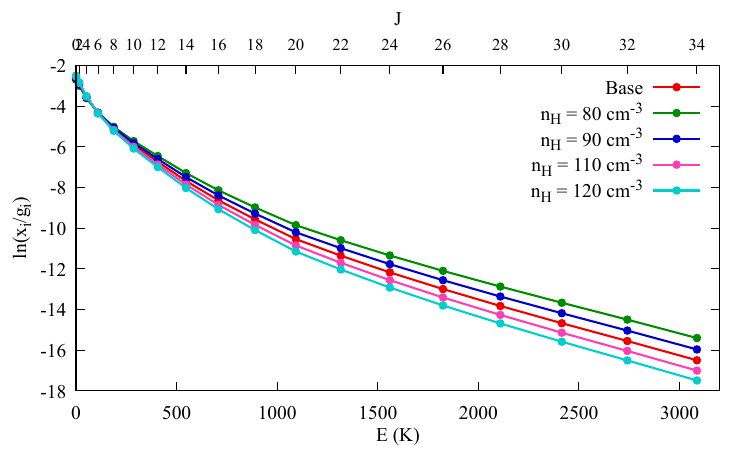}
\caption{Variation of $n_\mathrm{H}$
}
\label{Fig:Variations_n}
\end{figure}
\begin{figure}[h!]
\centering\includegraphics[width=1\columnwidth]{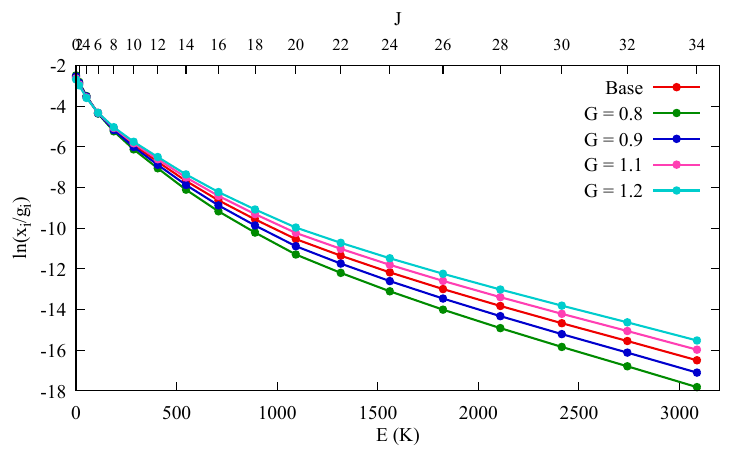}
\caption{Variation of $G$
}
\label{Fig:Variations_G}
\end{figure}

\end{appendix}

\end{document}